\documentclass[fleqn,usenatbib]{mnras}

\usepackage{newtxtext,newtxmath}

\usepackage[T1]{fontenc}
\usepackage{ae,aecompl}

\usepackage{graphicx}	
\usepackage{amsmath}	
\usepackage{amssymb}	

\title[Sulphur and COMs in infrared cold cores]{Sulphur-Bearing and Complex Organic Molecules in an Infrared Cold Core}

\author[Beaklini et al.]{Pedro P. B. Beaklini$^{1}$,\thanks{E-mail: pedro.beaklini@iag.usp.br}
Edgar Mendoza$^{1}$,
Carla M. Canelo$^{1}$,
Isabel Aleman$^{1,2}$,
\newauthor{Manuel Merello$^{1}$,
Shuo Kong$^{3}$,
Felipe Navarete$^{1}$,
Eduardo Janot-Pacheco$^{1}$,
}
\newauthor{Zulema Abraham$^{1}$,
Jacques R.D. L\'{e}pine$^{1}$,
Amaury A. de Almeida$^{1}$,
}
\newauthor{Am\^{a}ncio C.S. Fria\c{c}a$^{1}$}
\\
$^{1}$Instituto de Astronomia, Geof\'{\i}sica e Ci\^{e}ncias Atmosf\'{e}ricas, Universidade de S\~{a}o Paulo.\\Rua do Mat\~{a}o 1226, 05508-090, S\~{a}o Paulo/SP, Brazil.\\ 
$^{2}$UNIFEI, Instituto de F\'{i}sica e Qu\'{i}mica, Universidade Federal de Itajub\'{a}, Av. BPS 1303 Pinheirinho, 37500-903 Itajub\'{a}, MG, Brazil\\
$^{3}$Department of Astronomy, Yale University, New Haven, CT 06511, USA
}

\date{Accepted XXX. Received YYY; in original form ZZZ}

\pubyear{2019}

\begin{document}
\label{firstpage}
\pagerange{\pageref{firstpage}--\pageref{lastpage}}
\maketitle

\begin{abstract}
Since the start of ALMA observatory operation, new and important chemistry of infrared cold core was revealed. Molecular transitions at millimeter range are being used to identify and to characterize these sources. We have investigated the 231 GHz ALMA archive observations of the infrared dark cloud region C9, focusing on the brighter source that we called as IRDC-C9 Main. We report the existence of two sub-structures on the continuum map of this source: a compact bright spot with high chemistry diversity that we labelled as core, and a weaker and extended one, that we labelled as tail. In the core, we have identified lines of the molecules OCS(19-18), $^{13}$CS(5-4) and CH$_{3}$CH$_{2}$CN, several lines of CH$_{3}$CHO and the k-ladder emission of $^{13}$CH$_{3}$CN.We report two different temperature regions: while the rotation diagram of CH$_{3}$CHO indicates a temperature of 25 K, the  rotation diagram of $^{13}$CH$_{3}$CN indicates a warmer phase at temperature of $\sim450$K. In the tail, only the OCS(19-18) and $^{13}$CS(5-4) lines were detected. We used the  $Nautilus$ and the \textsc{Radex} codes to estimate the column densities and the abundances. The existence of hot gas in the core of IRDC-C9 Main suggests the presence of a protostar, which is not present in the tail.

\end{abstract}




\section{Introduction} 
\label{int}

Infrared dark clouds are complexes of sources in which many of them are starless cores. These sources are composed of dense gas at low-temperature and dust condensations, being optically thick at wavelength of $\approx 10\mu m$, and therefore they are identifed as dark compared with the Galactic IR background emission \citep{but09}. These regions could become the birth place of massive stars \citep{deW05,bat14}. Temperatures inside such clouds are of only a few kelvins and densities are around 10$^{5}$ cm$^{-3}$, with masses of $\approx$ 50 M\textsubscript{$\odot$} \citep{but09,but12,kon17,kon18}. Cold core sources display a rich and complex chemistry, revealed by the detection of many molecular species that provide important clues about the early stages of stellar evolution \citep{cas12}. Besides the low temperature,  outflows in dark regions have been already detected in these sources  through observations of ALMA (Atacama Large Millimeter Array), NOEMA (North Extended Millimeter Array), and SMA (Submillimeter Array) \citep{tan16,fen16,pil19}.  Normally, the molecule N$_2$D$^{+}$ is used as a main tracer of temperature and density of these cold cores  \citep{cra05,tan13,kon17}.

The emission of sulphur-bearing molecules can be used as a chemical clock for such regions. Massive pre-stellar cores collapse to form hot cores in short time scales. The abundances of SO, CS, OCS and chemical evolution in such environments provide a method to follow their evolution, even at low temperatures. \citep{cha97,van03,wak04,her09,esp14,li2015}.

The presence of complex organic molecules (COMs) in proto-stellar cores is well established \citep[e.g.][]{ell80,bla87,cec98,kua04,bot04,arc08,jab14,bro15,alma2015,jor16}. Different models were developed to explain their formation, either in the gas or in solid phase \citep[eg.][]{bro88,char92,has93,hor04,gar06,rua15}. In the past few years, the presence of COMs was detected in cold cores, indicating a non-thermal process for the formation of such species \citep[e.g.][]{bac16}. After methanol, first detected in 1985 \citep{mat85}, there are only a few other COM detected in dark cold regions. After the first detection by \citet{obe10} in the low-mass protostar B1-b, there are few works that we can mention: \citet{cer12} also in B1-b, \citet{bac12} and \citet{bac16b} in L1689B, \citet{vas14} and \citet{jim16} in L1544, and \citet{som18} in the Taurus Molecular Cloud 1. Because of the low temperature of these sources, the identification of weak COM emission lines from relatively cold objects/environments requires observations with large telescopes using long integration times.

Some attempts have been made to understand the formation processes of COMs in the cold environment. In all the cases, methanol is the precursor of more complex organic species. The abundances of methanol and OCS in such cold environments indicate that grain surface formation is very important \citep{gar07,loi12}. The chemistry in the dust ice mantles can be much richer and efficient than gas phase production at low-temperatures, with the existence of different mechanisms that may influence the molecular formation \citep{gar06, gar08,vas12,cha16,che18,kal18}. In particular, \citet{rua15} showed that a scenario which includes the Eley-Rideal mechanism seems to be necessary to explain all the observations, although the current calculations still underestimate the abundance of methanol in the gas phase. As a consequence, the inclusion of grain chemistry is important to explain the observations \citep{gar06}.

It is still unclear if the formation of sulphur-bearing molecules occurs via gas or solid phase and which is the sulphur reservoir. Observations of sulphur-bearing species abundances in comets and their connection with proto-stars advocate for a solid phase formation route \citep{boc00,dro18}. They can also provide constraints to be used in the grain chemistry models \citep{mar16}. The recent detection of the thioformyl radical (HCS) and its metastable isomer, as well as their relative abundances with respect to H$_{2}$CS indicate that the sulphur chemistry is far from being fully understood \citep{agu18}.


In the present work, we use the ALMA observations of a sample of infrared dark clouds obtained by \citet{kon17} to search for new molecular line emission and study the chemistry in starless cold cores. Is it possible to detect sulphur-bearing molecules and carbon molecular complex species in these cores? What can such emission reveal about temperature and chemical and physical evolution of the early stage of massive clump collapse? Is there a possibility that an eventual starless core presents a warming up phase only detected in high resolution observations? These are the questions that we want to address in this work.   

The high sensitivity of ALMA make this instrument ideal to detect such weak cold sources at high resolution. \citet{kon17} performed a survey at band 6 (centred at rest frequency of 231.32 GHz) of starless cores, searching in 32 clumps compiled from the sample of \citet{but12} with temperatures between 10-15~K. They have reported emission of N$_{2}$D$^{+}$, C$^{18}$O, DCO$^{+}$, CH$_{3}$OH, SiO, and DCN. Here we re-analyse the ALMA data for most of their sources and report the detection of several other species toward one of their infrared dark cloud complexes, the IRDC-C9 source. In the present paper, we follow the same nomenclature for the IRDCs used by \citet{but12} and \citet{kon17}.



The source IRDC-C9 is a complex of cold cores with 6 identified clumps of N$_{2}$D$^{+}$, including the brightest N$_{2}$D$^{+}$ flux among all sources of the sample observed by \citet{kon17}: IRDC-C9A. It is located at 5 kpc, and the coordinates of the system centre are:  $\alpha$ (J2000) $=$ 18$^{h}$42$^{m}$51.86$^{s}$ and $\delta$ (J2000) $=$ -03$^{o}$59${'}$22.3${''}$, with a local standard of rest velocity of 79 $\rm{km s^{-1}}$ \citep{but12,kon17}. In Galactic coordinates, it is at  l = 28.39950$^\circ$ and  b = 0.08217$^\circ$. Beyond the C9 complex, other 8 cold core complexes, identified as IRDC-C 1 to 8, were located nearby, in an area of 20 $\rm{arcmin^{2}}$.

The paper is organized as follows: we provide details on the data and the reduction procedure in Sect.~\ref{met}, we show the results of our data analysis in Sect. ~\ref{res}, the discussion of our results in Sect.~\ref{dis}, and conclusions are summarized in Sect.~\ref{conclusion}.

\begin{figure}
	\includegraphics[width=\columnwidth]{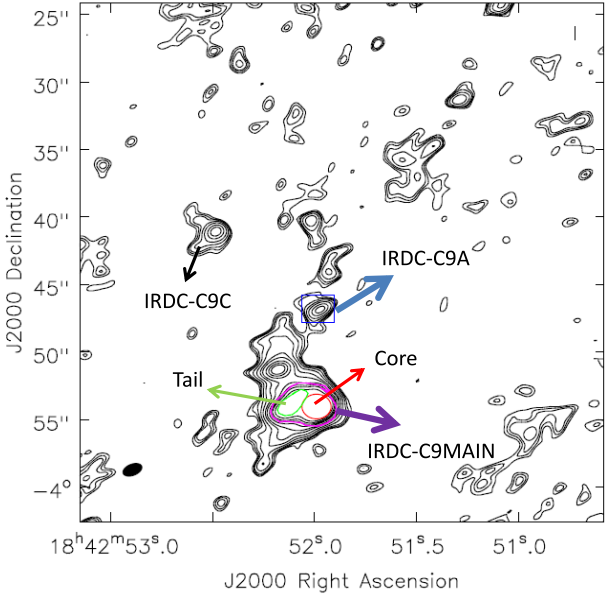}
    \caption{ IRDC-C9 region at 231.5 GHz.  The contours are placed at the same levels indicated in Fig.5 from  \citet{kon17}: contour levels of 0.22 mJy times 3, 4, 5, 7, 10, 15, 20, 30, 40, 50, 70. Both figures are not exactly the same because we perform our own self-calibration method and we are looking to another spw, however, it is easy to identify the sources. The N$_{2}$D$^{+}$ (3--2) core IRDC-C9A is indicated by a blue box, while IRDC-C9 Main is pointed by the magenta line contours, with the core in red and the tail in green. A detailed colour map of IRDC-C9 Main is shown in Fig. \ref{f1}. }

    \label{complex}
\end{figure}


\section{Observations and Data Analyses}
\label{met}
\subsection{Data Reduction}

The observations reported by \citet{kon17} provide an unprecedented survey of massive starless cores. The observations (project code 2013.1.00806.S, P.I. J. Tan) were carried out with ALMA band 6 (211-275 GHz), with the most compact configuration of the 12m array during the observation cycle 2. They searched for N$_2$D$^+$ emission in 32 infrared dark clouds (IRDCs), showing maps of the flux distribution for each line for all the regions and identifying many IRDCs sub-structures. More details on the observations can be found in \citet{kon17}. 

The data was divided into 4 different spectral windows: two very narrow bandwidth centred at N$_{2}$D$^{+}$ (rest frequency at $231.32$ GHz), and C$^{18}$O (rest frequency at $219.56$ GHz), another was split in four parts to detect other lines, and the last one, centred at 231.5~GHz, with a 2 GHz  bandwidth was used to obtain the continuum emission. As the ALMA spectral windows are divided into channels, this band also gives a low-resolution spectrum. We performed the data analysis using \textsc{CASA}\footnote{\url{https://casa.nrao.edu/}} (Common Astronomy Software Application) in this large spectral band, searching for any additional molecular lines. The spectral line analyses was carried out with the CLASS/GILDAS\footnote{\url{https://www.iram.fr/IRAMFR/GILDAS/}} and CASSIS\footnote{\url{http://cassis.irap.omp.eu/}} softwares. Both tools were used to estimate the physical conditions under Local and non-Local Thermodynamic Equilibrium, LTE and non-LTE assumptions, respectively.  Spectroscopic  databases such as Splatalogue,\footnote{\url{https://www.cv.nrao.edu/php/splat/}} CDMS,\footnote{\url{https://www.astro.uni-koeln.de/cdms}} and JPL\footnote{\url{https://spec.jpl.nasa.gov/}} were used in this work.

After following the same calibration procedure applied by \citet{kon17}, we used the standard Hogboom algorithm available in \textsc{CASA} to clean the image with Briggs weighting. Such procedure resulted in a 256$\times$256 pixels map with $0\farcs 05$  cell sizes. We obtained our final continuum band image after performing self-calibration, which was also applied to the ALMA data cube. The ALMA data cube is formed by 1920 channels, with a frequency bandwidth of $980$ KHz per channel (1.26 $\rm{km s^{-1}}$). The line spectra through this paper are presented at rest frequency. 

The original data of the project were divided into two tracks. We focused on the track containing the sources identified as A, B, C and E. From the list of total 13 dark cores, only the IRDC-C9 complex presented prominent line emission features in the observed frequency range and, therefore, it is the one reported here. The source we analysed is not the N$_{2}$D$^{+}$ (3--2) core IRDC-C9A, instead, we looked at the bright continuum source of IRDC-C9, which we labelled as IRDC-C9 Main. In Fig. \ref{complex} we marked the position of IRDC-C9A with a blue square, while IRDC-C9 Main is marked in magenta and it can be divided in two substructures in red and green.  IRDC-C9 Main is not coincident with any of the N$_{2}$D$^{+}$ sources previously reported.

\subsection{Modelling Methods}
\label{chem-model}

To obtain the temperature and column density that best explain the line emission we use two different approaches. For the molecules that have at least three detected lines, we use the rotation diagram as usual (e.g. \citealt{gol1999}). However, when it was not the case, we use the RADEX \footnote{\url{https://home.strw.leidenuniv.nl/~moldata/radex.html}}  code to estimate the physical conditions for both substructures. We applied the Markov Chain Monte Carlo method (MCMC) \citep{Foreman2013}\footnote{\url{http://dfm.io/emcee/current/}}, which is henceforth referred to as LTE/MCMC and RADEX/MCMC (non-LTE/MCMC). The MCMC method calculates the best solution by means of numerical chains in a $n$-dimensional parameter space defined by different free parameters, such as the column density, kinetic temperature, full width at half maximum (FWHM), line position, source angular size and volume density $n$(H$_2$). The best solution is calculated and achieved through stochastic processes to find the minimum value of the $\chi^2$ function.

RADEX requires collisional coefficients of the modelled molecules and an estimation of the column density of H$_2$ \citep{van07}. The  collisional coefficients were taken from the LAMDA (Leiden Atomic and Molecular Database).\footnote{\url{http://home.strw.leidenuniv.nl/~moldata/}} In particular, the collisional excitation mechanisms of OCS with species such as He and H$_2$ are discussed in \citet{Green1978} and in \citet{Lique2006}. For the volume density, we use the value of $n$(H$_2$)=4.6 $\times$ 10$^6$~cm$^{-3}$ obtained by \citet{kon17} for IRDC-C9A, supposing that is not too different in IRDC-C9 Main (see Fig. \ref{complex} to identify both sources ). 

For the chemistry of the source, we used the $Nautilus$ code \citep{nautilus}. $Nautilus$ is a useful tool to understand the grain and gas phase reactions of hot and cold cores \citep{sem10,reb14,rua15,nautilus}. The model predicts the time evolution of the chemical abundances for a given set of physical and chemical parameters. For the solid-state chemistry, it considers mantle and surface as chemically active, following the formalism of \citet{has93} and the experimental results of \citet{ghe15}. It can perform a three-phase (gas plus grain mantle and surface) time-dependant simulation of the chemistry in hot and cold cores, including chemical reactions in both gas and solid phases \citep{rua15}. 

The grain chemistry also considers the standard direct photodissociation by photons along with the photodissociation induced by secondary UV photons \citep{Prasad83}, which are effective processes on the surface and mantle of the grains.  Moreover, the  competition between reaction, diffusion and evaporation, suggested by \citet{chang07} and \citet{pauly11} is also implemented in the model. The diffusion energies of each species are computed as a fraction of their binding energies and we used the same proportion chosen by \citet{vid18} --  0.4 for the surface and 0.8 for the mantle \citep[for further details see][]{nautilus, vid18}.
\begin{figure}
	\includegraphics[width=\columnwidth]{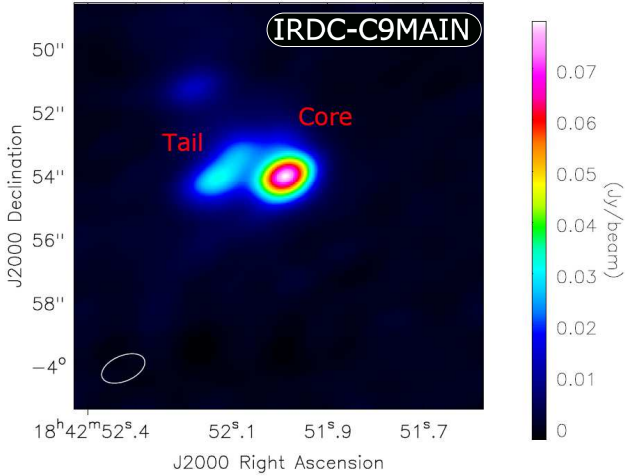}
    \caption{ Continuum map of IRDC-C9 Main, showing a core structure emission and a tail nearby. The flux is presented in colour scale in units of $Jy/beam$ The resolved beam is  $1.4 \times 0.7$ arcsec}
    
    
    
    \label{f1}
\end{figure}

\begin{figure*}
	\includegraphics[width=16cm]{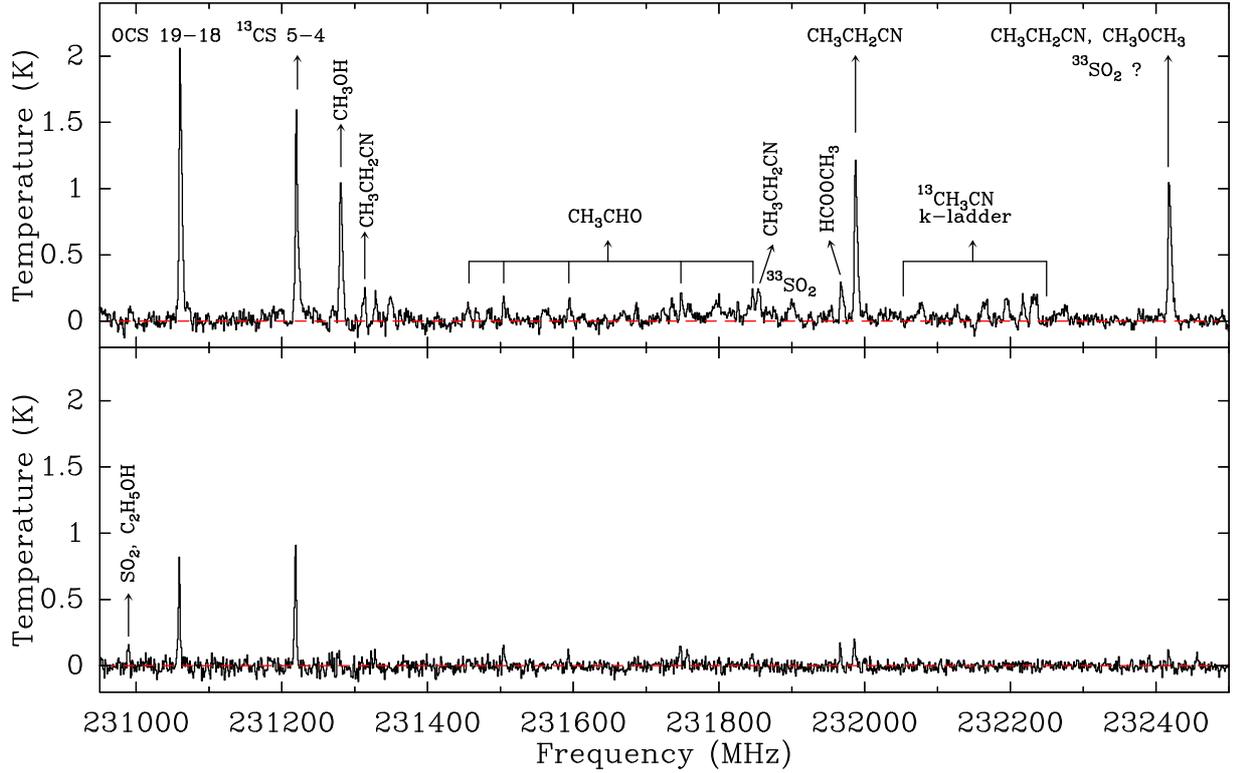}
	
    \caption{Spectra toward the core (top) and tail (bottom) of C9 Main showing the identification of the most intense lines. Spectral baselines were subtracted in the line-free segments; the red-dashed lines indicate the standard baseline.}

    
    \label{f2}
\end{figure*}

\section{Results}
\label{res}

\subsection{Continuum and line emission from IRDC-C9 Main}

We present the continuum emission of IRDC-C9 Main in Fig.~\ref{f1}, where the surface brightness colour scale has units of $\rm{Jy beam^{-1}}$. In this continuum map we can clearly see the presence of the distinct substructures associated with the  source: a bright and compact emission (referred as ''core'') and a more extended and weaker structure (refereed as ''tail''). The total flux density of the core is 9.2 $\times$ 10$^{-2}$ Jy, while the tail is  4.6 $\times$ 10$^{-2}$ Jy. Both structures are slightly larger than the beam, shown on the bottom left corner of the figure.   


In Fig ~\ref{f2} we show the spectra of the core and of the tail, integrated over the spatial distribution of each component. The rms of the spectrum in the core and in the tail are 44 and 36 mK, respectively.  In the core we detected lines of OCS (19--18), $^{13}$CS (5--4), methanol and CH$_{3}$CH$_{2}$CN (propionitrile). The fifth bright line, located at $232.41721$ GHz, has not been clearly identified; it could be CH$_{3}$OCH$_{3}$ (dimethyl ether), other propionitrile line or even $^{33}$SO$_{2}$. 


The detection of sulphur-bearing molecules in cold regions is not unusual. OCS has been identified in hot cores, indicating a core with high density, and in cold cores, as the pre-stellar core L1544   \citep{pal97,hat98,vit04,esp14,vid18,vas18}. Methanol is expected in cold cores since it is a precursor of COMs \citep{gar06}. Propionitrile has been recently detected in the hot core associated with the IRDC clump G34.43+00.24 MM. Its abundance relative  to methanol can be used to determine if it is a high or low mass pre-stellar core \citep{tak15,sak18}. A molecule that is likely originated from the methanol reaction is  dimethyl ether \citep{gar06,pee06,bro13} and it has been detected towards the cold cores L1689B and B1-b \citep{bac12,cer12}. However, we are not sure that the fifth peak (at 232.41179 GHz) is due the existence of dimethyl ether and we attributed it to CH$_{3}$CH$_{2}$CN, because we have detected other lines of this molecule. In this way, we can not rule out the possibility of contamination.

The spectrum of the tail, presented in the bottom part of Fig.~\ref{f2}, is different from the core spectrum. There are only OCS (19--18) and $^{13}$CS (5--4) as prominent lines, and the second one is brighter than the first one. Besides a weak emission of HCOOCH$_{3}$ and CH$_{3}$CH$_{2}$CN, all others molecular lines are not detected.

Comparing the core and tail flux densities, we observe that the OCS line reaches 2.0 K in the core, that is 2.5 times the line flux density in the tail. CS, instead, reaches 1.6 K in the core and 0.9 K in the tail. The methanol line in the core  has 0.9 K, while in the tail is only marginally detected. The propionitrile and the dimethyl ether temperatures are 1.2 and 1.0 K in the core; 0.2 K and 0.1 K in the tail, respectively. They have slightly different velocities since there is a small shift between the lines  (around 1 km s$^{-1}$) when we compare both spectra.

There is only one line that is detected in the tail that is not seen in the core, located at 230.97 GHz, 4.6 times the tail rms. It could be SO$_{2}$, but it is difficult to observe this transition in low temperature environments. Another possibility, could be the ethanol molecule (C$_{2}$H$_{5}$OH). In fact, the bright lines of the spectrum of C9 main seem to be very similar to those in Sgr B2. There, the main source presents a $^{13}$CS line brighter than OCS line, while the North source presents the OCS line brighter than the $^{13}$CS line, and only the main source shows the emission of C$_{2}$H$_{5}$OH \citep{num98}. These are the same characteristics of IRDC-C9 Main spectrum, but with C$_{2}$H$_{5}$OH in the fainter continuum source.  In Table~\ref{tab:lines}, we exhibit the list of molecules identified in C9 Main, including its tail (Fig.~\ref{f2})

\subsection{Molecules in the core}
\subsubsection{OCS and $^{13}$CS}
\label{sec:cs}

In the spectrum exhibited in the top panel of Fig~\ref{f2}, we indicated the observed line of OCS, which corresponds to the $J$=19--18 transition. Although it was the brightest line observed in the frequency range, we could not build the OCS rotational diagram due to the absence of (at least) two more detections. 
Alternatively, we ran the RADEX non-LTE code using the Markov Chain Monte Carlo algorithm to estimate the physical conditions of the region. 

\begin{figure}
\centering
	\includegraphics[width=8cm,keepaspectratio]{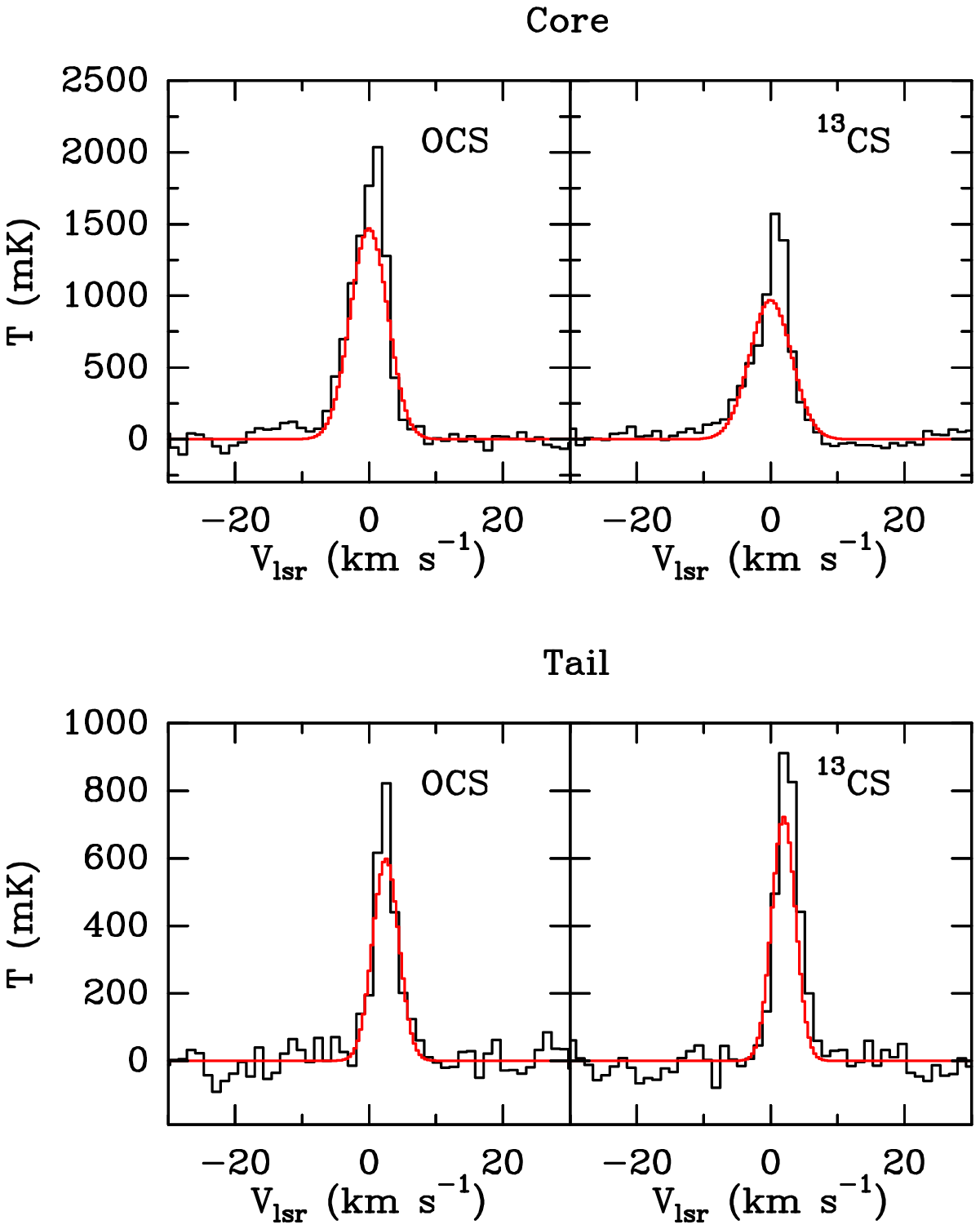}

    \caption{Spectral lines of OCS and $^{13}$CS detected in the core (top) and tail (bottom) of C9 Main. Black histogram lines correspond to the spectra and the red Gaussian curves indicate the synthetic lines obtained from the models assuming non-LTE conditions.}

    \label{panel-tail}
\end{figure}
\begin{figure*}

	\includegraphics[width=14cm]{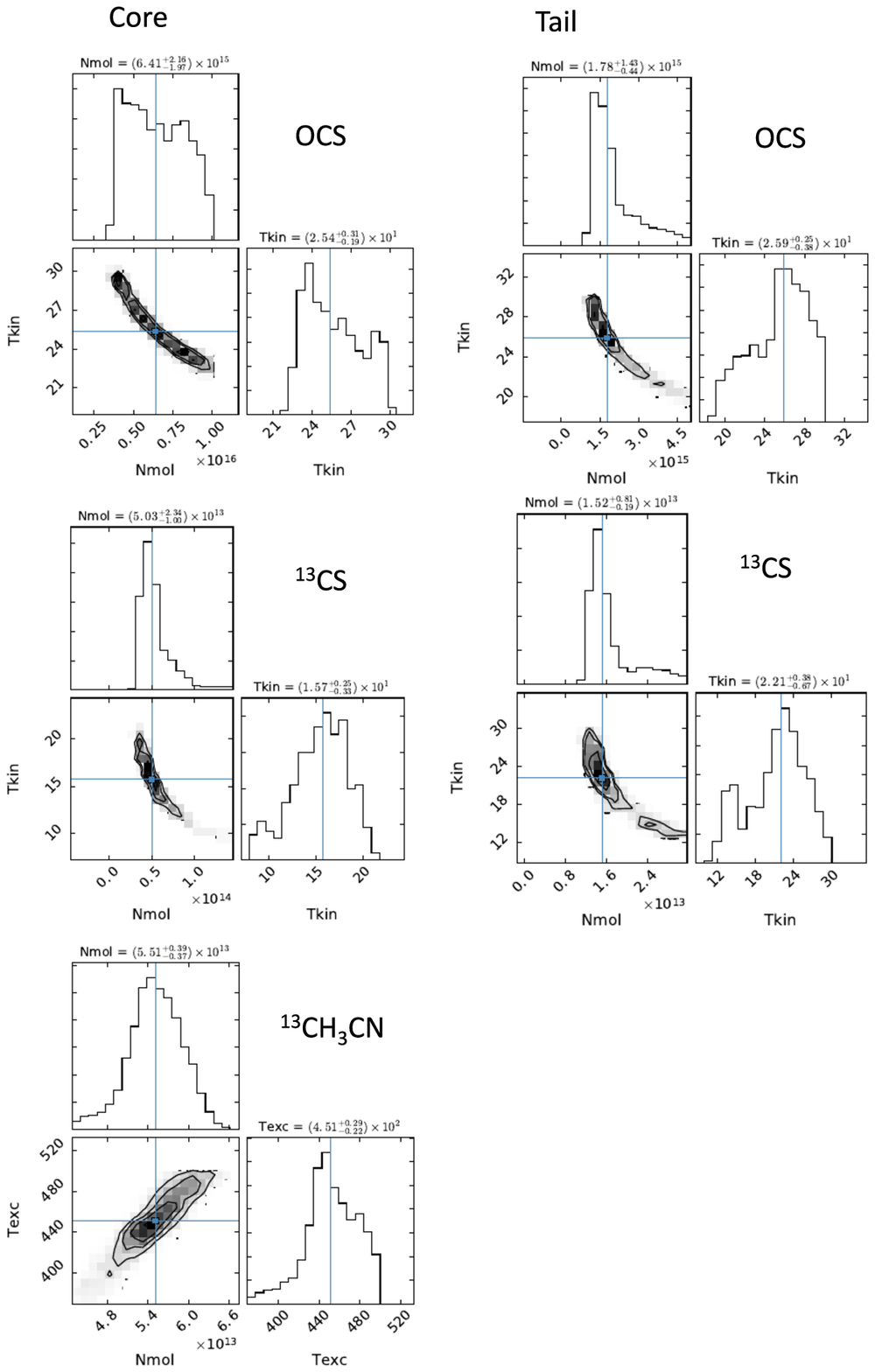}
    \caption {Corner diagrams derived for core and tail. For the core: the OCS (top), $^{13}$CS (middle),  $^{13}$CH$_3$CN (bottom). For the tail: OCS (top), $^{13}$CS (middle).  The solution for  $^{13}$CH$_3$CN was obtained via LTE/MCMC computations.  The solutions were obtained via RADEX/MCMC calculations, except for $^{13}$CH$_3$CN that was obtained via LTE/MCMC computations. The plots in each diagrams describe the projected histograms of the two parameters temperature (presented in Kelvin) and column density (presented in cm$^{-2}$). The contours of the corner diagrams (2D plots) represent the 1 and 2 $\sigma$ levels at the 39 and 86 per cent confidence levels.} 
    \label{diagrams}
\end{figure*}

The best fit yielded $N$(OCS)=6.41$\times$10$^{15}$~cm$^{-2}$ and $T_{kin}$=25~K for a convoluted source size of 2 $arcsec$, the reduced chi-square value was $\chi^2_{red}$=4.1. The synthetic line obtained from that solution is shown in the upper panel of Fig.~\ref{panel-tail}. In order to show the convergence of the free parameters column density and kinetic temperature, considering the applied RADEX/MCMC sampling, in the top left column of Fig.~\ref{diagrams} we presented the corner diagram for the achieved solution. The corner diagrams exhibit the highest probability distributions resulting from the individual histograms of the free parameters mentioned above.

We carried out a similar analysis to the \ $^{13}$CS (5--4) line. The best fit yielded $N$($^{13}$CS)~=~3.8~$\times$~10$^{13}$~cm$^{-2}$ and $T_{kin} =$ 20.7~K, the reduced chi-square value was $\chi^2_{red}$~=~3.22. We show in Fig.~\ref{panel-tail} the $^{13}$CS observed and simulated   spectra. \citet{li2015} reported the ratio C/$^{13}$C $\sim$ 45 in a sample of star-forming regions enriched by sulphur-bearing molecules. By applying such value to our observed emission of $^{13}$CS we estimated, approximately, that  OCS/CS $\sim$ 3.7, in agreement with several sources observed by \citet{li2015}. The corner diagram of $^{13}$CS is presented on the middle of the  left column of Fig. \ref{diagrams}.


\begin{figure}
\centering
	\includegraphics[width=7cm]{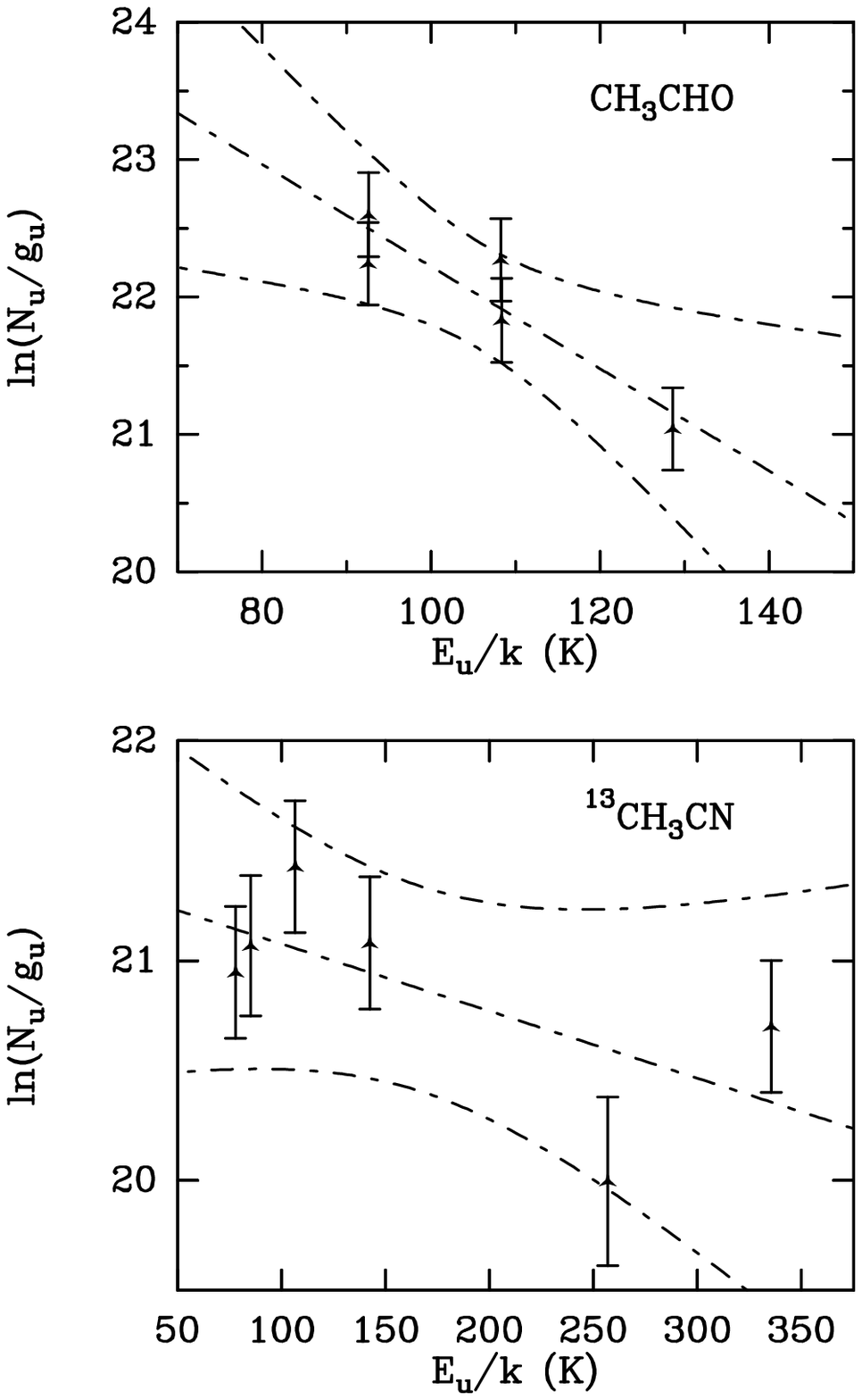}
    \caption{Rotational diagrams of: CH$_3$CHO (top), affected by multiplets at similar frequencies and upper energy values. $^{13}$CH$_3$CN (bottom). For both rotational diagrams, beam dilution corrections were applied assuming source size of up to 2 arcsec. The dashed-dotted lines indicate the linear fit including the confidence levels at 95 per cent. Error bars include calibration uncertainties of per cent.}

    \label{rd1}
\end{figure}


\subsubsection{CH$_3$CHO}
\label{sec:ch3cho}

We identified five lines of acetaldehyde (CH$_3$CHO) exhibiting $E_u$ values between $\sim$~80 and 140~K, being two of them affected by multiplets at similar frequencies and upper energy values.  In order to estimate its physical conditions, we carried out radiative analysis under LTE, assuming that the emission is optically thin and considering that the source fills the beam. By means of the rotational diagram analysis (e.g. \citealt{gol1999}), we estimate the CH$_3$CHO column density ($N$) and excitation temperature ($T_{exc}$) using
\begin{equation}
\ln \frac{N_u}{g_u} = \ln N - \ln Z - \frac{E_u}{kT_{exc}}
\end{equation}
where $N_u$, $g_u$ and $Z$ are the column density, degeneracy and partition function at the upper level $u$, respectively.

\begin{figure*}
\centering
	\includegraphics[width=16cm,keepaspectratio]{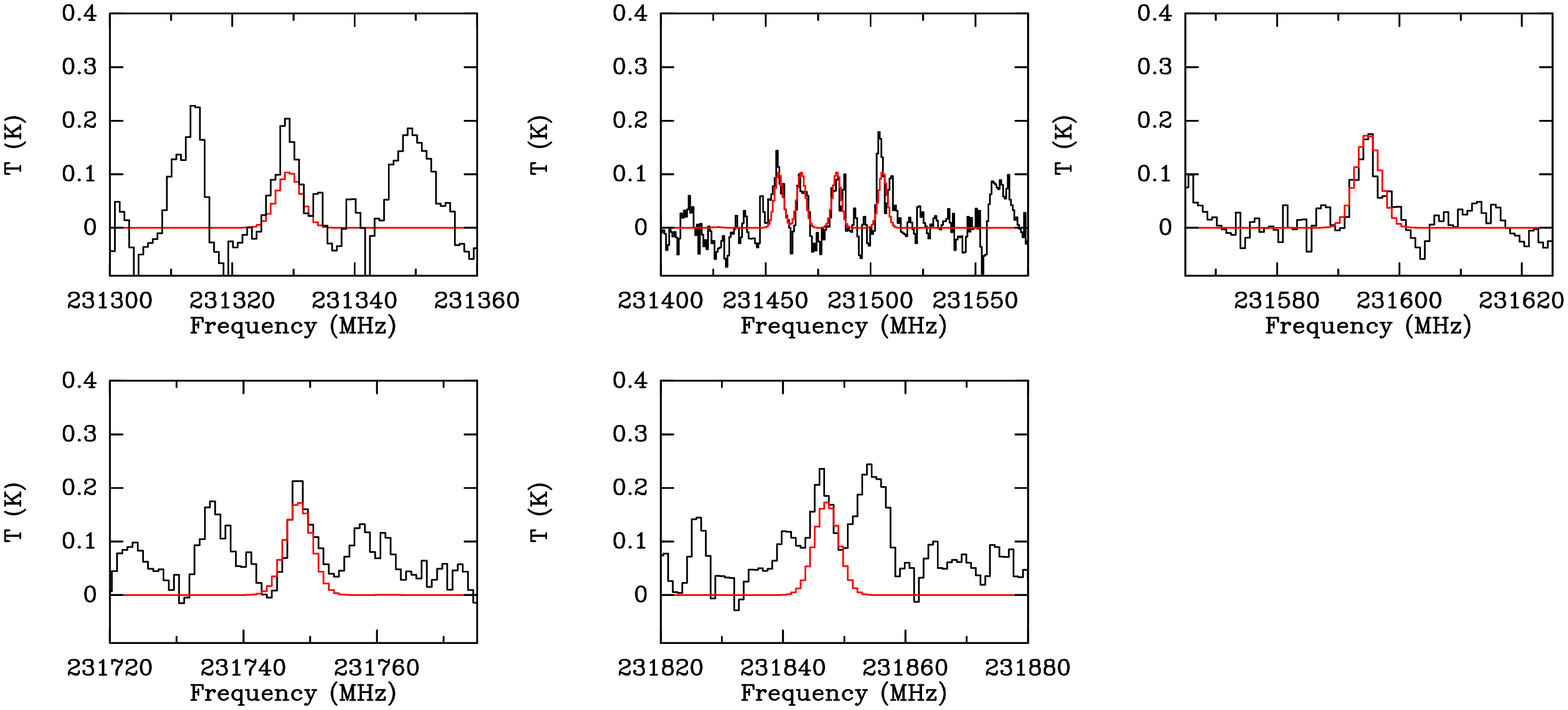}

    \caption{Spectral signatures of CH$_3$CHO identified in the  core of C9 Main. The black histogram lines represent the spectra and the red histogram lines are the CH$_3$CHO models obtained from the LTE solutions.}

    \label{panel}
\end{figure*}

The rotational diagram of the CH$_3$CHO molecule is presented in the top panel of Fig.~\ref{rd1}. From the linear least square regression, obtained after applying a beam dilution correction for a source size of 2$\arcsec$,  we obtained $N$(CH$_3$CHO) = (2.5 $\pm$ 1.5) $\times$ 10$^{14}$~cm$^{-2}$ and $T_{exc}$ = 27 $\pm$ 5~K. The synthetic spectra of the observed CH$_3$CHO lines are shown in Fig.~\ref{panel}.

\begin{table*}

\caption{Molecular lines identified towards IRDC-C9 Main. The line identification was carried out using the JPL and CDMS spectroscopic databases}

\label{tab:lines}
\begin{tabular}{lcccccccc}
\hline
Frequency 	&  	Molecule   & Transition    & $A_{ij}$      	& $E_{up}$ 	& Area   		& FWHM            		& Velocity & Intensity 			\\
(MHz)    	&     &                      	&  (s$^{-1}$)   	& (K)      	& (K km s$^{-1}$)   		&  (km s$^{-1}$) 	& (km s$^{-1}$)	& (K)       				\\
\hline

Core\\

231060.993	&	OCS	&	19-18	&	3.58$\times$10$^{-5}$	&	110.89	&	12.09	$\pm$	0.77	&	2.56	$\pm$	0.09	&	1.85	$\pm$	0.09	&	1.88	$\pm$	0.05	\\

231220.685	&	$^{13}$CS	&	5-4	&	2.51$\times$10$^{-4}$	&	33.29	&	8.22	$\pm$	0.71	&	2.32	$\pm$	0.11	&	0.82	$\pm$	0.11	&	1.42	$\pm$	0.06	\\

231281.11	&	CH$_3$OH	&	10-9	&	1.84$\times$10$^{-5}$	&	165.34	&	6.54	$\pm$	0.60	&	2.62	$\pm$	0.13	&	1.38	$\pm$	0.13	&	1.00	$\pm$	0.04	\\

231310.42	&	CH$_3$CH$_2$CN	&	24-23	&	1.04$\times$10$^{-3}$	&	153.42	&	1.03	$\pm$	2.79	&	2.07	$\pm$	2.07	&	-0.41	$\pm$	2.07	&	0.20	$\pm$	0.17	\\


231345.257	&	CH$_2$DCN	&	5$_{2,4}$-6$_{1,5}$	&	1.65$\times$10$^{-7}$	&	34.07	&	1.26	$\pm$	0.73	&	3.06	$\pm$	0.84	&	1.17	$\pm$	0.84	&	0.16	$\pm$	0.04	\\

231456.744	&	CH$_3$CHO$^a$	&	12$_{4,9,0}$-11$_{4,8,0}$	&	3.75$\times$10$^{-4}$	&	108.35	&	0.53	$\pm$	0.48	&	2.24	$\pm$	0.92	&	1.10	$\pm$	0.92	&	0.09	$\pm$	0.03	\\

231595.273	&	CH$_3$CHO$^a$	&	12$_{3,10,0}$--11$_{3,9,0}$	&	3.96$\times$10$^{-4}$	&	92.58	&	0.61	$\pm$	0.31	&	1.72	$\pm$	0.41	&	0.15	$\pm$	0.41	&	0.14	$\pm$	0.03	\\


231748.719	&	CH$_3$CHO$^a$	&	12$_{3,10,1}$--11$_{3,9,1}$ 	&	3.94 $\times$10$^{-4}$
	&	92.51	&	0.89	$\pm$	0.41	&	1.93	$\pm$	0.43	&	1.18	$\pm$	0.43	&	0.18	$\pm$	0.04	\\

231967.022	&	HCOOCH$_3$	&	20$_{9,11}$-20$_{8,12}$	&	1.56$\times$10$^{-5}$	&	177.83	&	1.60	$\pm$	3.31	&	2.61	$\pm$	2.10	&	2.31	$\pm$	2.10	&	0.24	$\pm$	0.17	\\

231990.41	&	CH$_3$CH$_2$CN	&	27$_{0,27}$-26$_{0,26}$	&	1.06$\times$10$^{-3}$	&	157.71	&	7.35	$\pm$	0.74	&	2.63	$\pm$	0.14	&	1.03	$\pm$	0.14	&	1.11	$\pm$	0.05	\\

232020.609	&	$^{13}$CH$_{3}$CN$^b$	&	13$_7$-12$_7$	&	7.65$\times$10$^{-4}$	&	428.41	&	--	&	--	&	--	&	$\leq 0.08 $	\\

232077.203	&	$^{13}$CH$_{3}$CN	&	13$_6$-12$_6$	&	8.48$\times$10$^{-4}$	&	335.513	&	0.84	$\pm$	0.37	&	3.34	$\pm$	0.71	&	-0.04	$\pm$	0.71	&	0.10	$\pm$	0.02	\\

232125.130	&	$^{13}$CH$_{3}$CN$^b$	&	13$_5$-12$_5$	&	9.19$\times$10$^{-4}$	&	256.873	&	--	&	--	&	--	&	$\leq 0.10$ 	\\

232164.3695	&	$^{13}$CH$_{3}$CN	&	13$_4$-12$_4$	&	9.76$\times$10$^{-4}$	&	192.51	&	1.05	$\pm$	0.73	&	3.66	$\pm$	1.19	&	1.28	$\pm$	1.19	&	0.11	$\pm$	0.03	\\

232194.906	&	$^{13}$CH$_{3}$CN	&	13$_3$-12$_3$	&	1.02$\times$10$^{-3}$	&	142.43	&	1.09	$\pm$	0.62	&	2.94	$\pm$	0.79	&	1.11	$\pm$	0.79	&	0.15	$\pm$	0.03	\\

232216.726	&	$^{13}$CH$_{3}$CN	&	13$_2$-12$_2$	&	1.05$\times$10$^{-3}$	&	106.65	&	0.65	$\pm$	0.65	&	1.73	$\pm$	0.77	&	1.38	$\pm$	0.77	&	0.15	$\pm$	0.06	\\

232229.822	&	$^{13}$CH$_{3}$CN$^c$	&	13$_1$-12$_1$	&	1.07$\times$10$^{-3}$	&	85.18	&	--	&	--	&	--	&	0.16	$\pm$	0.03	\\


232234.1886	&	$^{13}$CH$_{3}$CN$^c$	&	13$_0$-12$_0$	&	1.07$\times$10$^{-3}$	&	78.02	&	--	&	--	&	--	&	0.16	$\pm$	0.03	\\

232411.790	&	CH$_3$CH$_2$CN	&	21$_{4,17}$-22$_{0,22}$	&	3.64$\times$10$^{-4}$	&	896.78	&	7.24	$\pm$	0.53	&	2.99	$\pm$	0.12	&	2.05	$\pm$	0.12	&	0.97	$\pm$	0.03	\\

\hline

Tail\\


231060.993	&	OCS	&	19-18	&	3.58$\times$10$^{-5}$	&	110.89	&	3.05	$\pm$	0.33	&	1.55	$\pm$	0.09	&	0.83	$\pm$	0.09	&	0.78	$\pm$	0.04	\\

231220.685	&	$^{13}$CS	&	5-4	&	2.51$\times$10$^{-4}$	&	33.29 &	3.90	$\pm$	0.29	&	1.68	$\pm$	0.07	&	0.08	$\pm$	0.07	&	0.92	$\pm$	0.03	\\

231985.357	&	HCOOCH$_3$	&	20$_{9,12}$-20$_{8,13}$	&	1.57$\times$10$^{-5}$	&	177.83	&	0.75	$\pm$	0.32	&	1.89	$\pm$	0.40	&	1.11	$\pm$	0.40	&	0.16	$\pm$	0.03	\\

231990.409	&	CH$_3$CH$_2$CN	&	27$_{0,27}$-26$_{0,26}$	&	1.06$\times$10$^{-3}$	&	157.71	&	0.93	$\pm$	0.36	&	1.88	$\pm$	0.36	&	0.98	$\pm$	0.36	&	0.20	$\pm$	0.03	\\

\hline
\end{tabular}
\\
$^a$ We present only the CH$_3$CHO lines used to make the rotational diagram, since the others are on the limit of detection ($\leq$ 0.1 K).\\
$^b$Blanks represent weak transitions that we are not able to produce a good fitting. We use the peak value as a superior limit.\\
$^c$These line transitions are superposed, as shown in Figure \ref{kladder}.  \\
\end{table*}

\subsubsection{The nitriles CH$_3$CN and CH$_3$CH$_2$CN}
\label{sec:nitriles}

Methyl cyanide (CH$_3$CN) is known as one of the best thermometers of the Interstellar Medium (ISM). Specially, CH$_3$CN traces well the gas temperature in hot cores, since its different $k$-ladder spectral signatures appear usually thermalized at their physical conditions \citep{olm1993,fue14,beu2017}. As part of the spectral analysis carried out in this work, we  detected one of the CH$_3$CN isotopologues in C9 Main, $^{13}$CH$_3$CN. As we show in the top panel of Fig.~\ref{f2}, the identification of $^{13}$CH$_3$CN line was carried out within the frequency interval $\sim$~(232.000 -- 232.250)~GHz, since we observed the whole $k$-ladder structure for the $^{13}$CH$_3$CN (13--12) transition. Despite the weakness of the emission, with all the lines exhibiting intensities below 200~mK, the $k$-ladder appeared in its normal pattern containing eight spectral lines, whose intensities decrease as the $k$ number increases from $k=0$ to $k=7$, although some of them are only marginally detected.

\begin{figure*}
\centering
	\includegraphics[width=15cm,keepaspectratio]{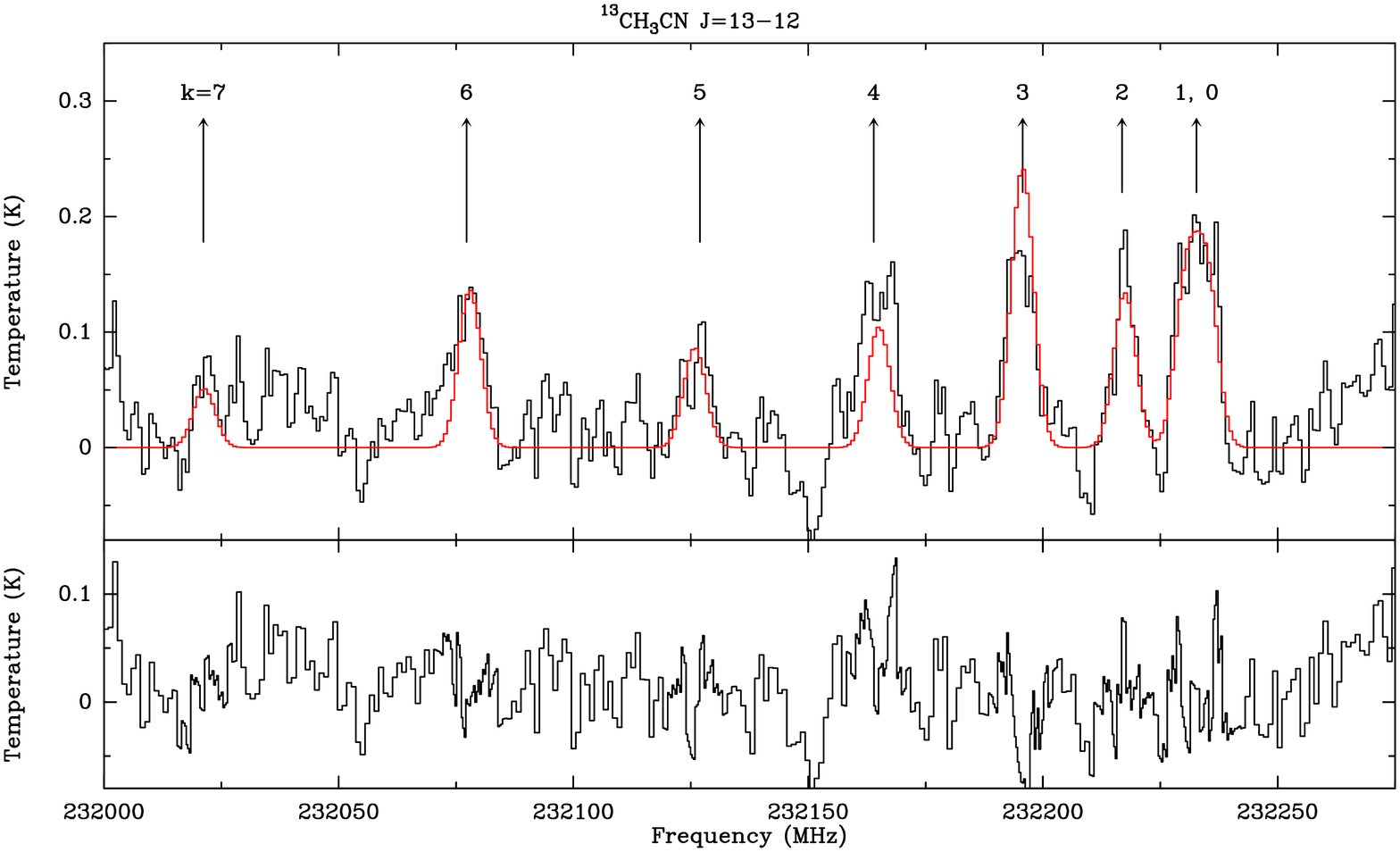}
    \caption{Spectral signatures of the $^{13}$CH$_3$CN $k$-ladder structure detected in the core of C9 Main. Black histogram represents the spectral lines identified by means of their $k$ numbers. The red solid line indicates the line synthetic spectrum from the LTE model. The inferior panel displays the residual spectrum, consisting of a subtraction between the observation and LTE model.}


    \label{kladder}
\end{figure*}

Due to the absence of collisional coefficients for $^{13}$CH$_3$CN, we used the population diagram method in order to estimate the physical conditions of the region; then excitation temperatures and column densities were derived from simple rotational diagrams. The LTE/MCMC computations of $^{13}$CH$_3$CN were carried out for temperatures up to 500 K. In the bottom panel of Fig~\ref{rd1} we showed the result, with a rotational diagram  yielding a straight line fit with an excitation temperature and column density of 460~K  and $N$($^{13}$CH$_3$CN) = 5.56 $\times$ 10$^{13}$~cm$^{-2}$, respectively. From the resulting LTE solution, in Fig.~\ref{kladder} the spectral model is displayed against the k-ladder structure of the $^{13}$CH$_3$CN transition (13--12). Alternatively, we also computed a LTE model using the MCMC algorithm with the column density and excitation temperature as free parameters. In agreement with the rotational diagram, we obtained a similar solution with $T_{exc} \simeq$ 450~K, as can be seen in the corner diagram on bottom left of Fig. \ref{diagrams} 

Due to the weakness and absence of more lines for a much trustworthy rotational diagram, we can only estimate the $^{13}$CH$_3$CN temperature, but it clearly reveals a higher excitation 
component in contrast with the gas temperatures of OCS and $^{13}$CS ($T_{kin}$ < 30~K); evidencing the existence of distinct physical condition within  C9 Main, which seems associated to pre-stellar sources at more evolved stages. In addition, the presence of N$_2$D$^+$ on the complex IRDC-C9 also supports a scenario where pre-stellar phases of star formation are occurring (e.g. \citealt{kon17}), since the chemical timescale is long enough to allow molecular  processes and detection of typical species of hot gas, such as CH$_3$CN. For instance, a fast warm-up phase physical model can consider 7.12 $\times$ 10$^4$ years for
a core to reach 450~K \citep{Garrod13}. 

As a tracer of warm ($T\simeq$500~K) regions, \citet{bel2014} observed the CH$_3$CN k-ladder structures with $J$ = 6--5, $J$ = 12--11, $J$ = 13--12 and $J$ = 14--13 towards the Orion Kleinmann-Low nebula; for the hot zone of IRc2, they estimated temperatures between $\sim$~400~K and 520~K. In very hot regions ($T\gg$500~K), \citet{beu2017} used ALMA observations at high angular resolution ($0\farcs 06$) to evidence possible fragmentation and disc formation in 
high-mass star forming regions. From LTE analysis of the CH$_3$CN $J$=37--36 k-ladder structure, observed around 690~GHz, they estimated excitation temperatures of up to 1300~K, whose emission is associated with the hotter gaseous disc surface layer. \citet{Moscadelli2018} also found high temperature regimes related to CH$_3$CN and CH$_3$CH$_2$CN towards the star forming region G24.78+0.08. In particular, they found that $^{13}$CH$_3$CN traces temperature values between 300--400 K.
    
In order to better constrain the $^{13}$CH$_3$CN excitation conditions, new observations of the main isotopologue CH$_3$CN J=13--12 could provide means to discuss aspects as the CH$_3$CN/$^{13}$CH$_3$CN abundance ratios, opacity of the emission, H$_2$ densities and the kinetic temperature related to the emission.

We also have identified several lines of CH$_3$CH$_2$CN in the core of IRDC-C9 Main. Although its presence in cold and young regions is debated, the detection of CH$_3$CN suggests a fast and efficient chemistry to produce ethyl cyanide.

\subsection{Molecules in the tail}

The continuum map presented in Fig~\ref{f1} reveals the existence of an extended emitting region of $\sim$~2\arcsec on the eastern direction of IRDC-C9 Main. Although we called this region the IRDC-C9 \lq\lq tail\rq\rq, it is not clear yet if the region is bounded to the core or it is an independent source. 

Similar to what was observed in the core, the strongest lines in the tail spectrum are OCS (19--18) and $^{13}$CS (5--4). They are, however, weaker than the core emission (see Fig.~\ref{f2}). Unfortunately, they can only provide a rough estimation of column density and temperature, because we only count with a single line for each of them. The best calculation ($\chi^2_{red}$=2.40) yielded $N$(OCS) $\simeq$ 1.8 $\times$ 10$^{15}$ cm$^{-2}$ and $T_{kin} \simeq$ 26~K.  Such values are similar to the physical conditions we inferred for the core. The non-LTE model of OCS is compared to the observed spectrum in the bottom panel of Fig.~\ref{panel-tail}. In the right column of Fig.~\ref{diagrams}, we show the OCS and $^{13}$CS corner diagram with the statistical solutions for the column density and kinetic temperature parameters.  This result is in agreement with the intensities if we apply the C/$^{13}$C ratio commented above.

Contrary to what was observed toward the core, in the tail, the $^{13}$CS line is brighter than the OCS line.  From statistical equilibrium calculations involving collisional coefficients between $^{13}$CS and H$_2$, we estimated that $N$($^{13}$CS) $\simeq$ 1.5 $\times$ 10$^{13}$cm$^{-2}$ and $T_{kin}$ $\simeq$ 22 K toward the tail. The synthetic lines are exhibited in the bottom panel of Fig.~\ref{panel-tail}.

\begin{figure}
\centering
	\includegraphics[width=8cm]{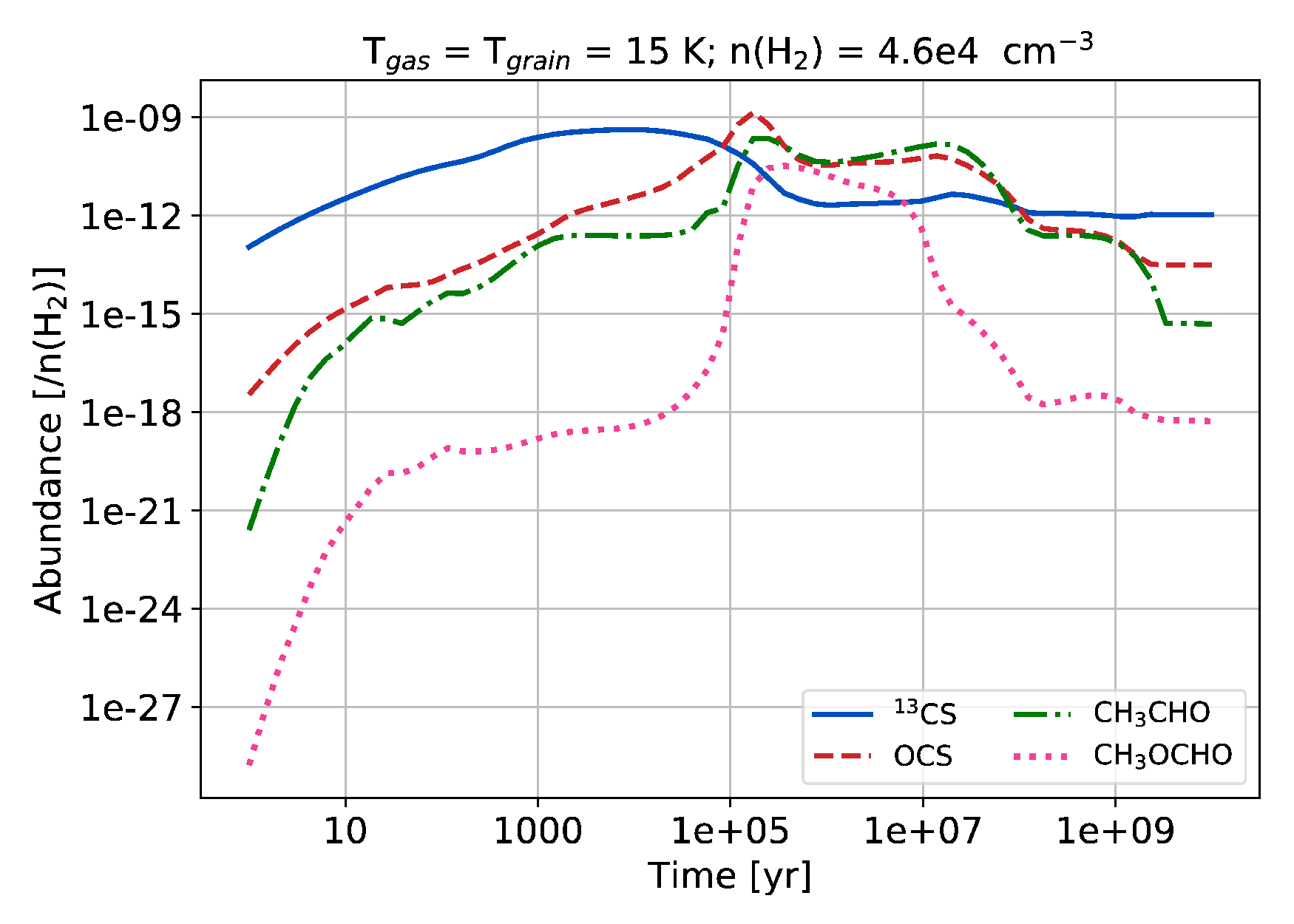}
    \includegraphics[width=8cm]{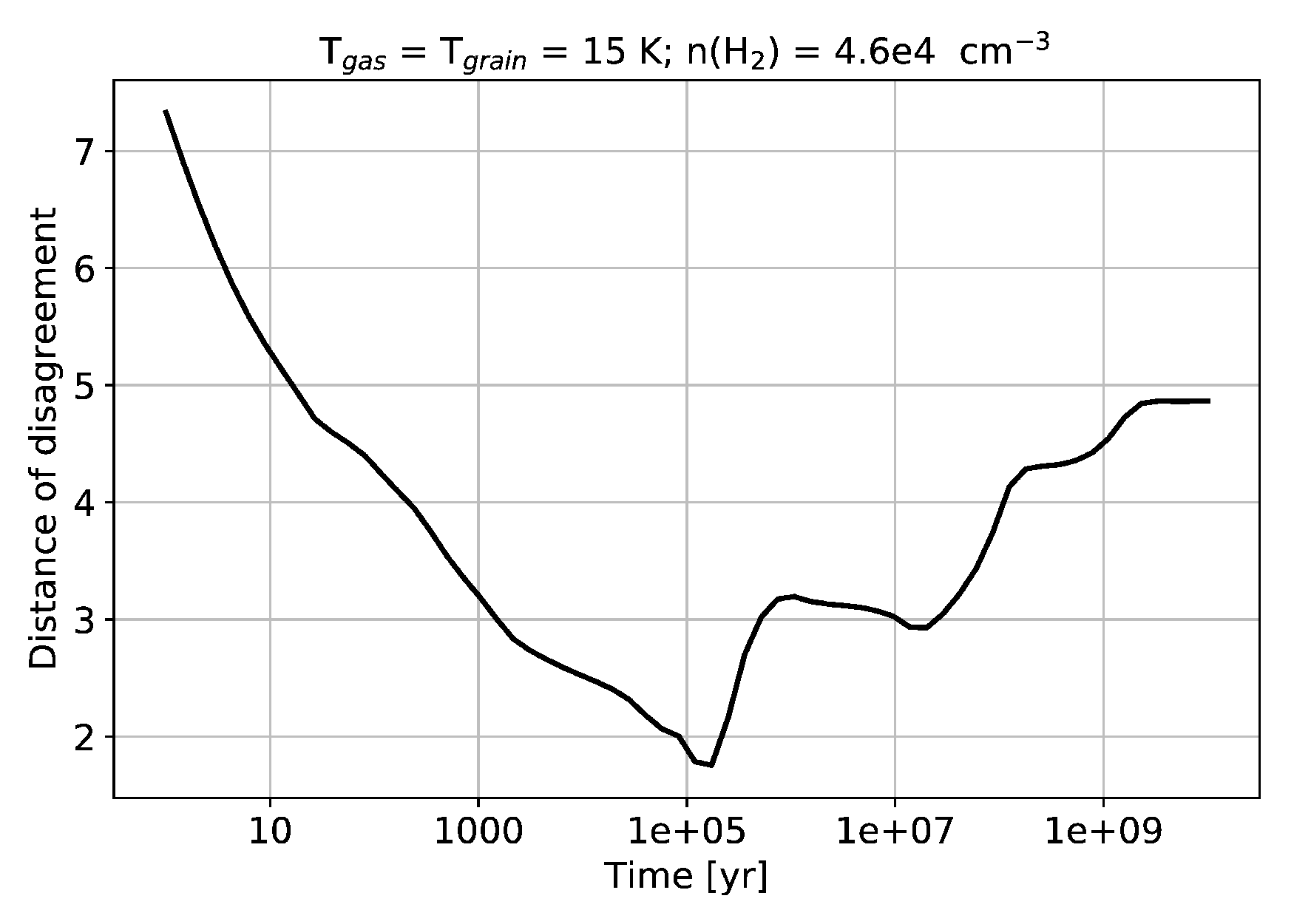}

    \caption{\textit{Top:} Time evolution of the abundances simulated with $Nautilus$. We assumed C/$^{13}$C = 45, and gas and grain temperatures of 25~K, and density of 10$^4$~cm$^{-3}$. \textit{Bottom:} Distance of disagreement \citep{wak06}, where the minimum indicates an estimation to the chemical age of $1.7 \times 10^{5} $years}
    \label{naut}
\end{figure}

\section{Discussion}
\label{dis}

\subsection{Molecular Abundances}
\label{DA}

From  the column densities of OCS, $^{13}$CS and CH$_3$CHO, we obtain their abundances estimating the column density of H$_{2}$ from the continuum flux density. We  used the equation derived by \citet{mot98} taking into account the beam size of 1.5 arcsec, using the dust opacity of 0.01 cm$^2$ g$^{-1}$, typical for pre-stellar cores, and a dust temperature of 15 K. We  obtained $N$(H$_2$)= 1.9 $\times$ 10$^{23}$~cm$^{-2}$ and  we present the abundances on Table \ref{abs}. Estimating the source size from the ALMA synthesised beam, we obtain the volume density $n$(H$_2$) $\approx$ 10$^6$~cm$^{-3}$,compatible with the value we used in the previous section.  

The comparison between the measured abundances for IRDC-C9 Main with the abundances of the sources L134 and TMC-1 \citep{fec15} shows compatible values for CH$_3$CHO and $^{13}$CS, considering C/$^{13}$C = 45.  However, we obtained higher abundance values for OCS. We have not estimated the abundance  of  $^{13}$CH$_3$CN, since the rotation diagram of this molecule shows a temperature of 450K and  the estimated H$_2$ column density was calculate for the cold region.  

Regarding OCS, as occurred for $^{13}$CS, we have detected only one line. The Radex/MCMC method considers quantities such as temperature and density. The column densities we have obtained for these molecules on the previous section are valid for a temperature of 25 K. If there is a hot region, it is also possible that it is contributing to the OCS line, increasing its emission. This explains why the OCS line presents a low brightness temperature with respect to the CS line in the tail, while in the core it is brighter. It could also explain the OCS overabundance.

There is another evidence of the existence of a warm region: the presence of the methanol line in the core. Methanol sublimates at 60K and we have not detected it in the tail, supporting the idea that only the core has a hot gas region inside. In fact, \citet{kon17} already suspected the existence of a protostar in this region, due to the detection of SiO in the IRDC-C9 complex. Using RADEX code, we can estimate the column density of methanol at two boundary temperatures; at 60 K and 250 K (typically high value). We obtained the column density of 1.0 $\times$ 10$^{16}$~cm$^{-2}$ for 60K and 4.0 $\times$ 10$^{15}$~cm$^{-2}$ for 250K, both with $T_{ex} \approx$ 30 K.    

To better understand the chemistry of the cold gas, we can use $Nautilus$ to verify the abundances and to estimate a chemical age. We included the molecules that we identified and that are presented in the $Nautilus$ database: CS, OCS, CH$_{3}$CHO and CH$_{3}$OCHO.  The code takes into account a complete network of reactions presented on the KIDA\footnote{\url{http://kida.obs.u-bordeaux1.fr/}} (KInetic Database for Astrochemistry) catalogue \citep{Wak15}. 

\begin{table}
	\caption{Abundances for IRDC-C9MAIN.}
	\label{abs}
    \centering
	\begin{tabular}{lrl}
		\hline
		\textbf{Element} & Core & Tail \\
		\hline
		 OCS        & 3(-8)  & 2(-8) \\
	     $^{13}$CS  & 9(-9)  & 8(-9) \\
		 CH$_3$CHO  & 8(-9)  & --    \\
		\hline
		\multicolumn{3}{l}{$^a$ Abundances given in the format}\\
		\multicolumn{3}{l}{a(b) representing a$\times$10$^{b}$.}\\
	\end{tabular}
\end{table}

\begin{table}
	\caption{Initial abundances assumed for the $Nautilus$ model.}
	\label{tab:abs}
    \centering
	\begin{tabular}{cccc}
		\hline
		\textbf{Element} & \textbf{n$_{i}$/n$_{H}$}$^{a}$ &  \textbf{Element} & \textbf{n$_{i}$/n$_{H}$}$^{a}$ \\
		\hline
		H$_2$ & 0.5  &  He & 9.0(-2)\\
		N  & 6.2(-5) &  O  & 2.4(-4) \\
		C+ & 1.7(-4) &  S+ & 1.5(-5) \\
		Fe+ & 3.0(-9)  & Si+ & 8.0(-9)\\
		Na+ & 2.0(-9)  & Mg+ & 7.0(-9)\\
		Cl+ & 1.0(-9)  & P+ & 2.0(-10)\\
        F & 6.7(-9)  & & \\
		\hline
		\multicolumn{4}{l}{$^a$ Abundances given in the format a(b) representing a$\times$10$^{b}$.}
	\end{tabular}
\end{table}

The simulations are zero-dimensional, i.e., total density, gas temperature, and other physical condition are uniform within the considered cloud and throughout the simulation time. There is no structure evolution in our model. We have assumed density of 4.6 $\times$10$^4$~cm$^{-3}$ and temperature of 15K for gas and grain.

The cloud initial elemental abundances were selected from \citet{vid18} and are displayed in Tab. \ref{tab:abs}. We assumed a standard cosmic ionization rate of 1.3~$\times$~10$^{-17}$~s$^{-1}$. We used a typical value for visual extinction of 10 mag. The code does not consider the isotopotologues $^{13}$CS and $^{13}$CH$_{3}$CN in the simulations. To calculate their column densities, we used the ratio of C/$^{13}$C $\sim$ 45, already mentioned, to estimate it from the corresponding most abundant isotopic molecule. 

The results of the model point to abundances similar to what we reported, as we present in Fig. \ref{naut}.  We have used the distance of disagreement method \citep{wak06}, a merit function that check for each time the predict abundances relative to the measured abundances. This indicates that the source has at least $10^{5}$ years old, the time necessary to reach the detected abundances, to estimate a chemical age of $1.7 \times 10^{5} $ years.  We did not modeled the abundances of the other molecules because they come from the hot gas in the core. 

\begin{figure}
\centering
	\includegraphics[width=6cm]{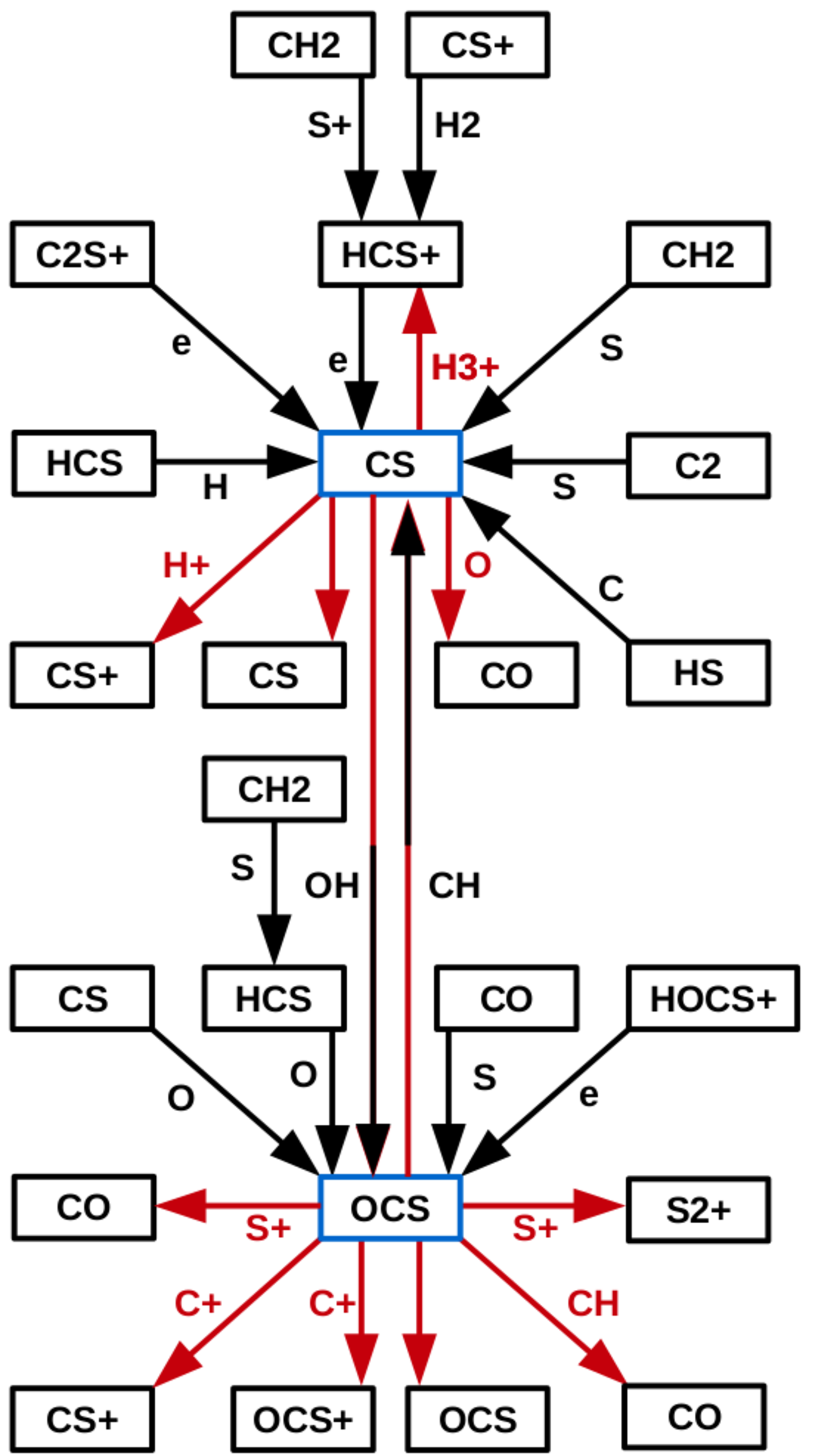}    
    \caption{CS and OCS chemical network, showing the main pathways of formation and destruction of the S-bearing molecules for a model with gas temperature of 25~K and grain temperature of 11.5~K. The black arrows represent the formation reactions and the red arrows are the destruction reactions of CS and OCS. A few other reactions contribute irrelevantly to the production or destruction of OCS and CS in this timescale of 10$^{3}$ to 10$^{4}$ years. }
    
    \label{fig:S-diag}
\end{figure}

In Fig. \ref{fig:S-diag} we present the reaction diagram for CS and OCS. It indicates that both suphur-bearing molecules are connected. The black arrows represent the formation reactions and the red arrows are the destruction reactions of CS and OCS.

\begin{figure}
\centering
    \includegraphics[width=8cm]{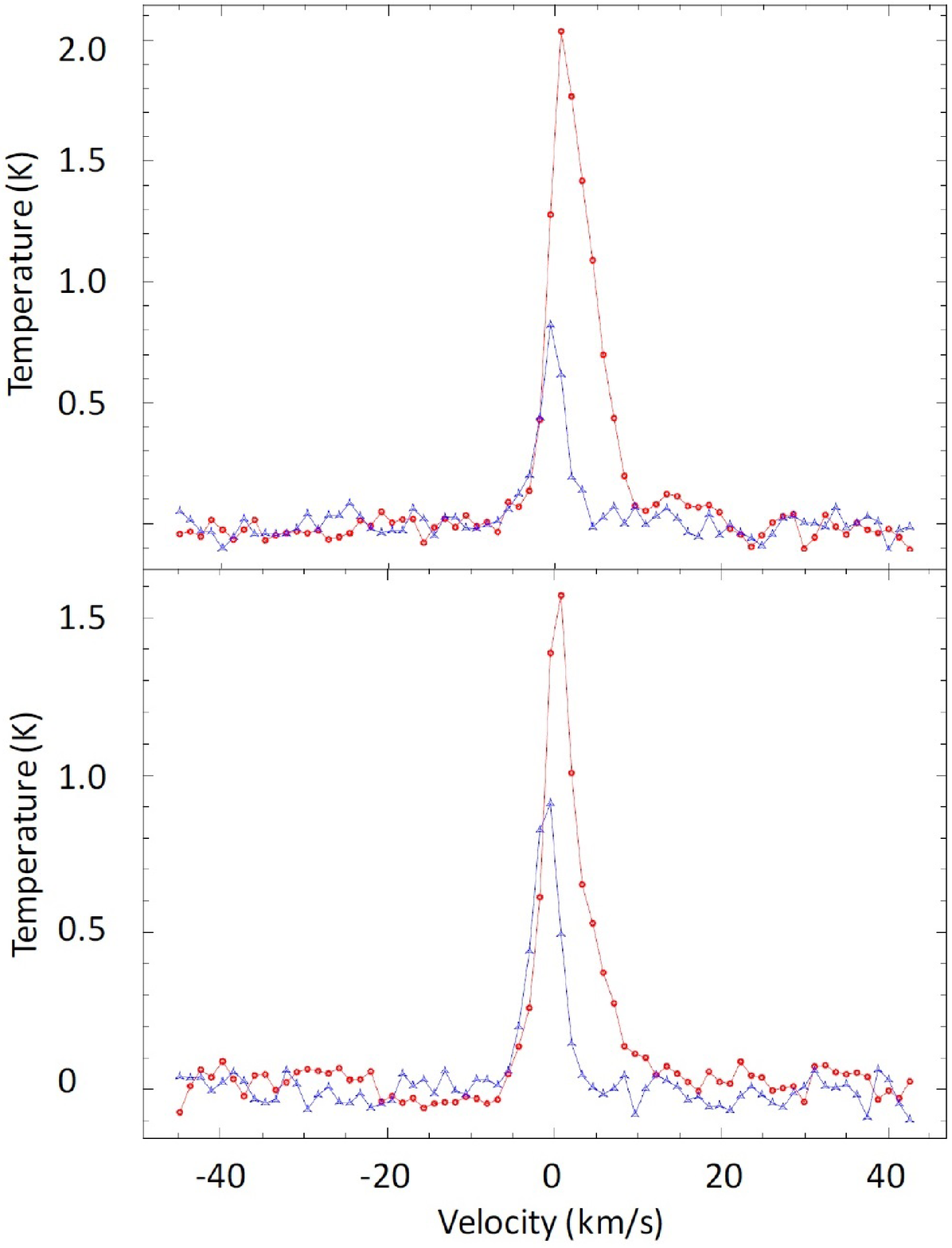} 
    \caption{OCS (on top) and $^{13}$CS (on bottom) line superposed core (red, circle) and tail(blue, triangles). Both lines do not have the same velocity between core and tail, and the OCS and the core have already the highest value. We attributed the slight asymmetry on the $^{13}$CS line at core due the opacity effects of an infalling gas.   }
    \label{fig:vel}
\end{figure}

\subsection{Core and tail characteristics}

The analysis of the IRDC-C9 complex using ALMA Band 6 data revealed a main bright source formed by two substructures, each of them with different morphological and spectral signatures.  The core continuum emission is brighter than the tail, with symmetric shape and size slightly larger than the beam size. The tail exhibits an asymmetric extended emission.

For cold dark clouds, the Spectral Energy Distribution (SED) can be modelled using two components, identified as a black body and grey body \citep{beu10}.At 231 GHz, the emission from the dust dominates the SED, and on cold cores, a single gray body component is required for fitting their SED \citep{beu10}. We can estimate the mass of core and tail using the grey body equation, which is similar to that used to compute de $H_2$ column density and is dependent on the mass and on the absorption coefficient \citep{elia10}. Again, we use an absorption coefficient at 230 GHz of $0.01 $cm$^2 g^{-1}$ \citep{pre93} and a temperature of 15 K. We obtain a mass of 57 M\textsubscript{$\odot$} for the core and 28 M\textsubscript{$\odot$} for the tail. If we suppose the continuum is a emitted by the hot gas, at 100 K, the mass should be only 3M\textsubscript{$\odot$}. This evidences that, indeed, the dust surrounded is in a cold temperature regime. 

The main physical difference between the substructures is the existence of the hot gas in the core. We can assume that the core is in a later evolutionary stage since a compact source already started to increase its temperature, while the tail has not evidence of a formed protostar (at least, not yet). This explains the absence of some lines in the tail. Furthermore, the emission of OCS is indicative of the existence of hot regions that originated in the gas surrounding the protostar \citep{guz14}.


Both core and tail are in a contraction process. The analysis of several starless cores indicated that a line shifted to the blue can be evidence of gas infalling process \citep{lee11}. Numerical simulations indicate that the infalling process would lead to an asymmetry in the blue and red wing of the line, but this asymmetry almost disappears in optically thin transitions\citep{sta10}. It is possible that we do not have enough spectral resolution to detect the two peak profile for the OCS line at core and tail. The slight asymmetry that we can see in Figure \ref{fig:vel} in the core could be evidence of such effect.

Another difference between core and tail is the slightly different velocities  between the lines, as we show in Figure \ref{fig:vel} and on Table \ref{tab:lines}. The OCS line for the core is at velocity 1.85 $\pm$ 0.09 km s$^{-1}$, while for the tail is at 0.82 $\pm$ 0.11 km s$^{-1}$. The $^{13}$CS lies at 0.83 $\pm$ 0.09 km s$^{-1}$ in the core and at 0.08 $\pm$ 0.03 km.s$^{-1}$ in the tail. The lines in the core are also broader than in the tail. This could be consequence of the outflow, as reported by the SiO emission by \citet{kon17}. The hot gas is presented only in the core, which increases its velocity and the FWHM of the line. In both structures, the OCS line has a higher velocity than the $^{13}$CS line. 
 

In the core, we detect the hot and the cold gas inside the ALMA beam, and then, we can see the spectral lines for both regions. The ALMA synthesized beam is very small (1.4 $\times$0.7 arcsec, see Fig.~\ref{f1}), and at the distance of 5 kpc \citep{but12}, it can resolve structures of $7 \times 10^{3}$ AU. It is reasonable to have such a temperature gradient at this scale. 
 
 We can check if the hot gas in the innermost region of IRDC-C9 Main could be detected at another wavelength. In fact, on the APEX survey of the galaxy, the ATLASGAL catalogue (APEX Telescope Large Area Survey of the GALaxy, \citealt{atlasgal}), the source AGAL028.398+00.081 at 0.87 mm seems to be correlated with IRDC-C9 Main, with galactic coordinates are l = 28.39950$^\circ$ and  b = 0.08217$^\circ$. It also shows a $^{13}$CO(3-2) line with $1.0$ K detected by the James Clark Maxwell telescope \citep{dem13}. At higher frequencies (11 and 22 $\mu m$), the continuum maps obtained from $Spitzer$-$WISE$ (Wide-field Infrared Survey Explorer) observations, show emission towards the southeast of AGAL028.398+00.081 , not detected at 0.87 mm \citep{con13}, that could be IRDC-C9 Main, detected now with high resolution observations using ALMA at 1.3 mm.

Regarding the chemistry of the core and the tail, we can argue that the difference of the line detection in each structure could be an indicative that the chemistry of the ice mantle is more relevant than expected for this object. In fact, the k-ladder of $^{13}$CH$_{3}$CN, molecule that is also detected in comets \citep{mard16}, is an indicative of the solid phase formation. When the temperature increases, the sublimation starts, and the detection occurs. We did not detect it in the tail because it has not been evaporated yet.

This can also be an explanation for the CH$_3$CH$_2$CN detection, which is brighter in the core than in the tail. Our interpretation is that the evaporation has just started in the tail, while in the core, most of this molecule is already in the gas phase. Another evidence of the chemistry in ice mantles is the presence of CH$_{2}$DCN because the deuterium molecules are difficult to form in the gas phase. In fact, \citet{kon17} reported the detection of  N$_{2}$D$^{+}$ and the DCN in IRDC-C9A, indicating the importance of the deuterium to the chemistry of the C9 complex.

We should mention also the existence of a line only marginally detected in the tail that could be SO$_{2}$ or C$_{2}$H$_{5}$OH. Normally, the sulphur molecules SO, OCS and $^{13}$CS are used as chemical clocks \citep{her09,esp14,li2015}. However, we did not find any chemical reactions that led to destroying SO$_2$ producing OCS. Based on the spectrum of Sgr B2 \citep{num98},  it is more reasonable to attribute this line to the molecule C$_{2}$H$_{5}$OH, but this remains as an open question.

\section{Conclusions}
\label{conclusion}

We investigated the data of \citet{kon17} on the ALMA archive that have provided an unprecedented list of identified starless cores. From this catalogue, we report the analysis of an exceptional molecular region on the infra-red dark cloud named IRDC-C9.  Other starless cores presented on the \citet{kon17} were analysed: IRDC-A,-B,-E,-H, but only IRDC-C9 presents prominent lines that are not a common detection even with the ALMA beam size.      

It is expected that interferometric observations of starless cores would be able to detect molecular line emission better than single dish observations. Since high-resolution observations mean a smaller beam, the dilution of each line flux is smaller in arrays than in single antennas.  The bright source of the infrared dark cloud complex of C9, IRDC-C9 Main, exhibits two substructures: a bright core and an extended tail. The bright core could host a protostar. The difference between velocities and FWHM between lines in the core and tail could point to a faint outflow in the core. 

We detected sulphur bearing molecules and complex organic molecules in  IRDC-C9 Main. The brightest source on the continuum image at 231.5 GHz presents two regions: a compact and rich molecular core with hot and cold gas and an extended tail. We detected OCS and $^{13}$CS in both structures, but we did not find hydrocarbons in the tail. The core continuum emission is brighter than the tail emission, revealing a mass of 57 M\textsubscript{$\odot$} for the core and 28 M\textsubscript{$\odot$} for the tail.

We suggest that the core is in a later stage of evolution. The contraction process of the innermost region is already at warming up phase, leading to the evaporation of some molecules, while the tail is in an initial stage of the collapse.  This is corroborated by the OCS line being more intense in the core since the OCS is typically found in gas around newly formed sources. In both cases,  the chemical age is at least 10$^5$ years, time necessary to reproduce the observed abundances.







\section*{Acknowledgements}

We would like to thank Dr. Bertrand Lefloch (IPAG/Universite Grenoble Alpes) for comments and discussions. We would like to thank  Dr. Jonathan C. Tan (University of Florida) for discussion and suggestions on the data analysis. We would like to thank  Dr. Jose Cernicharo (CSIC/Madrid) for discussion during the COSPAR Capacity Building Workshop: Advanced School on Infrared and Sub-millimeter Astrophysics - Data analysis of the Herschel, Spitzer, Planck and Akari missions and ALMA.
We are grateful to the S\~{a}o Paulo research agency FAPESP and for financial support (FAPESP Projects: 2014/07460-0, 2014/22095-6, 2017/23708-0, 2017/18191-8). I.A. acknowledges the support of CAPES, Ministry of Education, Brazil, through a PNPD fellowship. This study was financed in part by the Coordena\c{c}\~ao de Aperfei\c{c}oamento de Pessoal de N\'ivel Superior - Brasil (CAPES) - Finance Code 001. This	paper	makes	use	of	the	following	ALMA	data: 2013.1.00806.S, obtained  from ALMA Archive.	ALMA	is	a	partnership	 of	ESO	(representing	its	member	states),	NSF	(USA)	and	NINS	(Japan),	together	with	NRC	(Canada),	NSC	and	
ASIAA	(Taiwan),	and	KASI	(Republic	of	Korea),	in	cooperation	with	the	Republic	of	Chile.	The	Joint	ALMA Observatory	is	operated	by	ESO,	AUI/NRAO	and	NAOJ.  This work has made use of computing facilities of the Laboratory of Astroinformatics (IAG/USP, NAT/Unicsul), whose purchase was made possible by the Brazilian agency FAPESP (grant 2009/54006-4) and the INCT-A.



\bibliographystyle{mnras}
\label{lastpage}
\end{document}